\documentclass[12pt]{iopart}

\usepackage{iopams}  
\usepackage{graphicx,epic,eepic}
\def\d{{\mathrm{d}}}
\def\r{{\mathbf{r}}}

\newcommand{\betrag}[1]{\left\vert#1\right\vert}

\def\d{{\mathrm{d}}}

\def\det{{\mathrm{det}}}
\def\tr{{\mathrm{tr}}}

\def\Box{\kern0.5pt{\lower0.1pt\vbox{\hrule height.5pt
      width 6.8pt \hbox{\vrule width.5pt height6pt
        \kern6pt \vrule width.3pt} \hrule height.3pt
      width 6.8pt} }\kern1.5pt}

\begin{document}
\title[Analogue model  for quantum gravity phenomenology]{Analogue model for quantum gravity phenomenology}
\author{Silke Weinfurtner}
\address{
School of Mathematics, Statistics,  and Computer Science,\\
Victoria University of Wellington,\\ 
PO Box 600, Wellington, \\
New Zealand}
\ead{silke.weinfurtner@mcs.vuw.ac.nz}
\author{Stefano Liberati }
\address{International School for Advanced Studies, \\
Via Beirut 2-4, 34013 Trieste, Italy, \\ 
and INFN, Trieste}
\ead{liberati@sissa.it}
\author{Matt Visser}
\address{
School of Mathematics, Statistics,  and Computer Science,\\
Victoria University of Wellington,\\ 
PO Box 600, Wellington, \\
New Zealand}
\ead{matt.visser@mcs.vuw.ac.nz}
\date{26 October 2005; Revised ? ? 2005; \LaTeX-ed \today}
\begin{abstract}
So called ``analogue models'' use condensed matter systems (typically hydrodynamic) to set up an ``effective metric'' and to model curved-space quantum field theory in a physical system where all the microscopic degrees of freedom are well understood. Known analogue models typically lead to massless minimally coupled scalar fields. We present an extended ``analogue space-time'' programme by investigating a condensed-matter system --- in and beyond the hydrodynamic limit --- that is in principle capable of simulating the massive Klein--Gordon equation in curved spacetime. Since many elementary particles have mass, this is an essential step in building realistic analogue models, and an essential first step towards simulating quantum gravity phenomenology. Specifically, we consider the class of two-component BECs subject to laser-induced transitions between the components,  and we show that this model is an example for Lorentz invariance violation due to ultraviolet physics.    
Furthermore our model suggests constraints on quantum gravity phenomenology in terms of the ``naturalness problem'' and ``universality issue''.
\end{abstract}
\maketitle

\section{Introduction an motivation}

The purpose of quantum gravity phenomenology (QGP) is to analyze the physical consequences arising from various models of quantum gravity (QG)~\cite{LIV,Jacobson:2002hd}. One hope for obtaining an experimental grasp on QG is the generic prediction arising in many (but not all) models that  discrete space-time at the Planck scale, $M_{\mathrm{Pl}} = 1.2 \times 10^{19} \; \mathrm{GeV/c^2}$, typically induces low-energy violations of  Lorentz invariance (LI).
The breakdown of LI will manifest in a modification of the dispersion relation. We investigate Lorentz invariance violations (LIV) in the boost subgroup, leading to an expansion of the  dispersion relation in momentum-dependent terms,
\begin{equation} \label{eq:mod-disp}
E^2=
m^2\;c^4+p^2\;c^2+c^4\left\{\eta_1\, M_{\rm Pl}\, p/c+\eta_2\,p^2/c^2+\sum_{n\geq3}
\eta_n\,\frac{(p/c)^{n}}{M^{n-2}_{\rm Pl}}\right\} \, ,
\end{equation}
where both the quantity $p/(M_{\mathrm{Pl}}\;c)$ and the coefficients $\eta_{n}$ are dimensionless. \footnote{The particular inertial frame for these dispersion relations is generally given by the Cosmological Microwave Background (CMB) frame.}. \\
%
The present paper focuses on non-renormalizable effective field theory (EFT) with ultraviolet modifications in the dispersion relation \cite{jlm-ann}. There are two aspects of this model theory that are interesting to look at.
The so called \textit{naturalness problem} is correlated with the appearance of $\eta_{1}$ and $\eta_{2}$ in the dispersion relation (\ref{eq:mod-disp}). The low-order corrections do not appear to be  Planck suppressed, and would therefore always be dominant and in disagreement with observation.  In order to avoid this problem it is necessary to introduce a second scaling parameter, well below the Planck scale, which dominates at low momentum and \emph{only} affects the first and second order-terms. Such a scenario is not well justified within an EFT framework; in other words there is no natural suppression of the low-order modifications in these models.
A less problematic question is the so called \textit{universality issue}, addressing whether the LIV is particle-dependent or not. For a ``universal'' LIV the coefficients $\eta_{n}$ are the same for all types of particles. 

In order to contribute to this debate, we have chosen a rather unconventional path: We investigate the energy dependent behavior of sound waves in a 2-component Bose--Einstein condensate. Building on the existence of so-called ``analogue models'' (AM) for minimally coupled massless fields in curved space-times~\cite{Unruh, unexpected, ergosphere, Garay, BEC1, AM, broken}, we show how to extend the AM to include massive particles \cite{2BEC}, and that the dispersion relation (now written in terms of frequency and wavenumber) is modified by Lorentz violating terms at the analogue Planck scale \cite{QGPAM}:
\begin{equation} \label{ext_disp_rel}
\omega^2 = \omega_0^2 + \left(1 + \eta_{2} \right) \, c^2 \; k^2 + \eta_{4} \, \left(\frac{\hbar}{M_{\mathrm{Pl}}} \right)^2  \; k^4 + \dots \; .
\end{equation}
We further calculate the dimensionless coefficients $\eta_{2}$ and $\eta_{4}$ for both massive and mass-less quasi-particles and discuss the naturalness problem and universality issue in our 2-component AM.
%
%

\section{Sound waves in 2-component BECs}

The basis for our AM is an ultra-cold dilute atomic gas of $N$ bosons, which exist in two single-particle states $\vert 1 \rangle$ and $\vert 2 \rangle$.    For example, we consider two different hyperfine states, $\vert F=1,m_{F}=1 \rangle$ and $\vert F=2,m_{F}=2 \rangle$ of $^{87}Rb$ \cite{jenkins,trippenbach}.  They have different total angular momenta $F$ and therefore slightly different energies. That permits us, from a theoretical point of view, to keep $m_{1} \neq m_{2}$, even if they are approximately equal (to about one part in $10^{16}$). At the assumed ultra-cold temperatures the atoms only interact via low-energy collisions and the 2-body atomic potential can be replaced by a contact potential. That leaves us with with three atom-atom coupling constants, $U_{11}$, $U_{22}$, and $U_{12}$, for the interactions within and between the two hyperfine states. For our purpose it is essential to include an additional laser field, that drives transition between the two single-particle states.~\footnote{A more detailed description on how to set up an external field driving the required transitions can be found in \cite{Bloch}.} The rotating frame Hamiltonian for our closed 2-component system is given by:~\footnote{In general, it is possible that the collisions drive coupling to other hyperfine states. Strictly speaking the system is not closed, but it is legitimate to neglect this effect~\cite{dressed}. }
\begin{eqnarray}
\hat H = \int \d \r \;
\Bigg\{  
&&\sum_{i = 1,2}  \left(-\hat \Psi_{i}^\dag \frac{\hbar^2 \nabla^2}{2 m_{i}} \hat \Psi_{i}  
+ \hat \Psi_i^\dag V_{ext,i} (\r) \hat \Psi_i \right) 
\nonumber 
\\
&&+ \frac{1}{2} \sum_ {i,j = 1,2} \left(
 U_{i j} \hat \Psi_i^\dag  \hat \Psi_j^\dag  \hat \Psi_i  \hat \Psi_j  
 + \lambda  \hat \Psi_i^\dag (\mathbf{\sigma}_{x})_{i j}  \hat \Psi_j
\right)
\Bigg\} \, ,
\end{eqnarray}
with the transition energy $\lambda = \hbar \, \omega_{\mathrm{Rabi}}$ containing the effective Rabi frequency between the two hyperfine states. Here $\hat \Psi_i(\r)$ and $\hat \Psi_i^{\dag}(\r)$ are the usual boson field annihilation and creation operators for a single-particle state at position $\r$, and $\mathbf{\sigma}_x$ is the usual Pauli matrix.
For temperatures at or below the critical BEC temperature, almost all atoms occupy the spatial modes $\Psi_1(\r)$ and $\Psi_2(\r)$. The mean-field description for these modes,
\begin{equation}  \label{2GPE} 
 i \, \hbar \, \partial_{t} \Psi_{i} = \left[
   -\frac{\hbar^2}{2\,m_{i}} \nabla^2 + V_{i}-\mu_{i} + U_{ii}
   \, \betrag{\Psi_{i}}^2 + U_{ij} \betrag{\Psi_{j}}^2
   \right] \Psi_{i} + \lambda \, \Psi_{j} \, , 
\end{equation}
are a pair of coupled Gross--Pitaevskii equations (GPE):  $(i,j)\rightarrow (1,2)$ or  $(i,j)\rightarrow (2,1)$.

In order to use the above 2-component BEC as an AM, we have to investigate small perturbations (sound waves) in the condensate cloud.~\footnote{The perturbations have to small compared to the overall size of the condensate could, so the system remains in equilibrium.}  The excitation spectrum is obtained by linearizing around some background densities $\rho_{i0}$ and phases $\theta_{i0}$, using:
\begin{equation}
\Psi_{i}= \sqrt{\rho_{i0}+ \varepsilon \, \rho_{i1} }\,
e^{i(\theta_{i0}+ \varepsilon \, \theta_{i1} )}
\quad\hbox{for}\quad i=1,2 \, .
\end{equation}
A tedious calculation \cite{2BEC,QGPAM} shows that it is convenient to introduce the following $2 \times 2$ matrices: An effective coupling matrix,
\begin{equation}
\hat{\Xi}=\Xi+\hat X, 
\end{equation}
where
\begin{equation}
\Xi \equiv [\Xi]_{ij} =\frac{1}{\hbar}
\tilde{U}_{ij} = \frac{1}{\hbar} \left( U_{ij}-(-1)^{i+j} \, {\lambda\sqrt{\rho_{10}\rho_{20}}\over2} {1\over {\rho_{i0}\rho_{j0}} } \right)
\end{equation} 
and
\begin{equation}
\hat X \equiv  [\hat X]_{ij}= -{\hbar\over2} \, \delta_{ij} \, \frac{ \hat Q_{i1}}{m_i} 
= -{\hbar\over4}  \frac{ \delta_{ij} }{ m_i\;\rho_{i0}} 
= - [X]_{ij} \; \nabla^2 \, .
\end{equation}
Without transitions between the two hyperfine states, where $\lambda=0$, $\Xi$ only contains the coupling constants $[\Xi]_{ij} \rightarrow U_{ij}/\hbar$. While $\Xi$ is then independent of the energy of the perturbations, $\hat X$ plays a more significant role with the energy of perturbation. For low energetic perturbations, in the so-called hydrodynamic approximation, $\hat X$ can be neglected, $\hat X \rightarrow 0$, and $\hat \Xi \rightarrow \Xi$. 

Besides the interaction matrix, we also introduce a transition matrix,
\begin{equation}
\Lambda\equiv [\Lambda]_{ij}= -\frac{2\lambda\;\sqrt{\rho_{i0}\,\rho_{j0}} }{\hbar} \, (-1)^{i+j} \, ,
\end{equation}
and a mass-density matrix,
\begin{equation}
D\equiv [D]_{ij} = \hbar \, \delta_{ij} \,  \frac{ \rho_{i0} } {m_{i} } \, .
\end{equation} 

The final step is to define two column vectors,  ${\bar{\theta}} = [\theta_{11},\theta_{21}]^T$
and  ${\bar{\rho}} = [\rho_{11},\rho_{21}]^T$. We then obtain two compact equations for the perturbation in
the phases and densities,
\begin{eqnarray} \label{thetavecdot}
\dot{\bar{\theta}}&=&  -\,\Xi \; \bar{\rho} - \vec v_0  \cdot \nabla \bar{\theta},
\\
\label{rhovecdot}
\dot{\bar{\rho}}&=& \, - \nabla \cdot \left( D \; \nabla \bar{\theta} +  \bar{\rho} \; \vec{v_0} \right) 
- \Lambda \;\bar{\theta} \, ,
\end{eqnarray} 
where the background velocity $\vec{v}_0$ is the same in both condensates. Now combine these two equations into one:
\begin{equation} \label{phaseequation} \fl
\partial_{t} (\Xi^{-1} \; \dot{\bar{\theta}} ) =
 - \partial_{t} \left(\Xi^{-1} \; \vec v_0 \cdot \nabla \bar{\theta} \right) 
 - \nabla (\vec v_0 \; \Xi^{-1} \; \dot{\bar{\theta}} )  
 + \nabla \cdot \left[ \left(D - \vec v_0 \; \Xi^{-1} \; \vec v_0 \right) \nabla \bar{\theta} \, \right] 
 + \Lambda \; \bar{\theta}.  
\end{equation}
In the next section we show how this equation is analogous to a minimally coupled scalar field in
curved space-time. 


\section{Emergent space-time in the hydrodynamic limit}

Instead of keeping the analysis general \cite{2BEC,Ralf}, we now focus on the special case when $\Xi$ is independent of space and time, and on the hydrodynamic limit where $\hat X \rightarrow 0$. Then defining
\begin{equation} \label{tildephase}
\tilde\theta = \Xi^{-1/2}\; \bar\theta,
\end{equation}
equation (\ref{phaseequation}) simplifies to 
\begin{equation} \label{phaseequation2}
\fl 
\partial_{t}^2\tilde{\theta} =
 - \partial_{t} \left(\mathbf{I} \; \vec v_0 \cdot \nabla \tilde{\theta} \right) 
 - \nabla    \cdot   \left(\vec v_0 \; \mathbf{I} \; \dot{\tilde{\theta}} \right)  
 + \nabla \cdot \left[ \left(C_0^2 - \vec v_0 \; \mathbf{I} \; \vec v_0  \right) \nabla \tilde{\theta} \right] 
 + \Omega^2 \; \tilde{\theta},
\end{equation}
where
\begin{equation} \label{Omega2}
 C_{0}^2 = \Xi^{1/2}\; D \;\Xi^{1/2} \, ; 
\qquad \hbox{and} \qquad
\Omega^2 =   \Xi^{1/2} \;\Lambda\; \Xi^{1/2}.
\end{equation} 
Both $C_{0}^2$ and $\Omega^2$ are symmetric matrices. If $ [C_{0}^2, \; \Omega^2] = 0$,
which is equivalent to the matrix equation $ D \; \Xi \; \Lambda = \Lambda \; \Xi \; D$, then they have common eigenvectors.
Assuming (for the time being) simultaneous diagonalizability,  decomposition onto the eigenstates of the system results in a pair of independent Klein--Gordon equations
\begin{equation} \label{KGE}
\frac{1}{\sqrt{-g_{\mathrm{I/II}}}}
\partial_{a} \left\{   \sqrt{-g_{\mathrm{I/II}}} \; (g_{\mathrm{I/II}})^{ab} \; 
\partial_{b} \tilde{\theta}_{\mathrm{I/II}} \right\} + 
\omega_{\mathrm{I/II}}^2 \;
\tilde{\theta}_{\mathrm{I/II}} = 0 \; ,
\end{equation}
where the ``acoustic metrics'' are given by
\begin{equation} \label{metric}
(g_{\mathrm{I/II}})_{ab}=\left( \frac{\rho_{\mathrm{I/II}}}{c_{\mathrm{I/II}}} \right)^{2/(d-1)}
\left[
\begin{array}{ccc}
-\left( c_{\mathrm{I/II}}^2-v_0^2 \right)       &|& -\vec{v_0}^{\,T} \\
\hline
-\vec{v_0}  &|& \mathbf{I}_{d\times d}
\end{array}
\right] \, ,
\end{equation}
and where the overall conformal factor depends on the spatial dimension $d$. 
The metric components depend only on the background velocity $\vec{v}_{0}$, the background densities $\rho_{i0}$,  and the speed of sound of the two eigenmodes, which is given by
\begin{equation}
\label{e:csq-Xi}
c_{\mathrm{I/II}}^2 = 
\frac{\tr[C_0^2] \pm \sqrt{\tr[C_0^2]^2 - 4 \, \det[C_0^2]}}{2} \,.
\end{equation}
Considering the line element obtained from the acoustic metric (\ref{metric}), it is clear that the speed of sound in the AM takes the role of the speed of light. \\
It is also possible to calculate the eigenfrequencies of the two phonon modes,
\begin{equation}
\omega_{\mathrm{I}}^2 = 0; 
\qquad
\omega_{\mathrm{II}}^2 = \tr[\Omega^2] \, .
\end{equation}
A zero/ non-zero eigenfrequency corresponds to a zero/ non-zero mass for the phonon mode. They both ``experience'' the same space-time if
\begin{equation}
\tr[C_0^2]^2 - 4 \det[C_0^2]=0 \, .
\end{equation} 
The fact that we have an AM representing both massive and massless particles is promising for QGP if we now extend the analysis to high-energy phonon modes so that $\hat X \neq 0$. For the following, we concentrate on flat Minkowski space-time, by setting $\vec{v}_0 =\vec{0}$.
%

\section{QGP beyond the hydrodynamic limit }
Starting from equation (\ref{phaseequation}) for a uniform condensate, we set the background velocity to zero, $\vec{v}_0 =\vec{0}$, but keep the quantum pressure term, $\hat X \neq 0$. The equation for the rotated phases 
\begin{equation} \label{tildephase2}
\tilde\theta = \hat \Xi^{-1/2}\; \bar\theta
\end{equation}
 in momentum space is then \cite{QGPAM}
\begin{equation}
\omega^2 {\tilde{\theta}}  = 
\left\{
\sqrt{\Xi+X\; k^2} \;\; [D\; k^2+\Lambda]\;\; \sqrt{\Xi+X\;k^2} 
\right\}\;  \tilde{\theta}  = H(k^2) \; \tilde{\theta} \, .
   \label{eq:new-disp-rel}
\end{equation}
Thus the perturbation spectrum must obey the generalized Fresnel equation:
\begin{equation}
\det\{ \omega^2 \;\mathbf{I} - H(k^2) \} =0 \, .
\end{equation}
That is, the dispersion relation for the phonon modes in a 2-component BEC is
\begin{equation}
\omega_{\mathrm{I/II}}^2 = { \hbox{tr}[H(k^2)] \pm \sqrt{
    \hbox{tr}[H(k^2)]^2 - 4\;  \det[H(k^2)] }\over 2},
\label{eq:tot-disp-rel}
\end{equation}
and a Taylor-series expansion around zero momentum gives
\begin{equation} \label{Taylor} \fl
\omega_{\mathrm{I/II}}^2 = 
\left. \omega^2_{\mathrm{I/II}} \right\vert_{k \rightarrow 0}  
+ \left. \frac{\d \omega_{\mathrm{I/II}}^2}{\d k^2} \right\vert_{k \rightarrow 0} \; k^2
+ \left. \frac{1}{2} \, \frac{\d^2 \omega_{\mathrm{I/II}}^2}{\d \left(k^2\right)^2} \right\vert_{k \rightarrow 0} \, \left(k^2\right)^2
+ \mathcal{O}\left[( k^2)^3 \right] \, .
\end{equation}
These are two dispersion relations in the desired form of equation (\ref{ext_disp_rel}). Below we will explicitly compute equation (\ref{Taylor}) up to the fourth order. The fact that $\omega_{\mathrm{I/II}}^2 = \omega_{\mathrm{I/II}}^2(k^2)$ only permits even powers in $k$, therefore the dispersion relation is invariant under parity. This is by no means a surprising result, because the GPE (\ref{2GPE}) is also invariant under parity.

We now define the symmetric matrices
\begin{equation} \label{C2}
C^2 = C_0^2 + \Delta C^2\,;  
\qquad \Delta C^2 = X^{1/2}\Lambda X^{1/2}  \, ;
 \end{equation}
\begin{equation} \label{Z2}
Z^2 =  2 X^{1/2} D X^{1/2} = {\hbar^2\over 2} M^{-2}.
\end{equation}
Note that all the relevant matrices (equations \eref{Omega2}, \eref{C2}, and \eref{Z2}) have been carefully symmetrized, and note the important distinction between $C_0^2$ and $C^2$.
Now define
\begin{equation}
c^2 = {1\over2}\tr[C^2] \, ,
\end{equation}
which approaches the speed of sound $c^2 \rightarrow c_0^2$, in the hydrodynamic limit $C^2 \rightarrow C_{0}^2$, (see equation (\ref{e:csq-Xi})).
The second and fourth order coefficients in the dispersion relations (\ref{Taylor}), for a detailed calculation see~\cite{QGPAM}, are:
\begin{eqnarray} 
\fl \label{omega_I}
\left.{\d\omega_{\mathrm{I/II}}^2\over\d k^2}\right|_{k\to0} 
&= c^2 \left[ 1\pm \left\{
 2 \tr[\Omega^2 C_0^2 ] - \tr[\Omega^2]\;\tr[C^2] \over
 \tr[C^2] \tr[\Omega^2]  \right\}
 \right]
= c^2 (1\pm \eta_2) \, ; \\ 
\fl
\left.{\d^2\omega_{\mathrm{I/II}}^2\over\d(k^2)^2}\right|_{k\to0} &=  {\textstyle 1\over \textstyle 2} \Bigg[
\tr[Z^2]  \pm \tr[Z^2] 
\pm 2 \frac{\tr[\Omega^2{C}^2_0]-\tr[\Omega^2]\;\tr[C_0^2]}{\tr[\Omega^2]}\tr[Y^2]
\nonumber\\
&
\qquad\qquad
\pm {\tr[C^2]^2- 4\det[C^2_0] \over \tr[\Omega^2] }
 \mp
{\tr[C^2]^2\over\tr[\Omega^2]} \eta_2^2
\Bigg] 
\\
&= 2 \eta_{4} \left( \frac{\hbar}{M_{pl}} \right)^2 \; .  \label{omega_II}
\end{eqnarray}

\section{Lorentz violations from UV physics}

In order to obtain LIV purely due to ultraviolet physics, we demand perfect special relativity for $\hat X \rightarrow 0$. In other words, we require all terms in the equations (\ref{omega_I}) and (\ref{omega_II}) which would otherwise remain in the hydrodynamic limit to be zero. The constraints we get are:
 \begin{eqnarray}
&&C1:\qquad \tr[C^2_0]^2-4\det[C^2_0]=0 \, ;\\
&&C2:\qquad 2\tr[\Omega^2 {C}^2_0]-\tr[\Omega^2]\tr[C^2_0]=0 \, .
\end{eqnarray}
Beyond the hydrodynamic limit, but imposing $C1$ and $C2$, the equations (\ref{omega_I}) and (\ref{omega_II}) simplify to:
\begin{eqnarray}
\left.{\d\omega_{\mathrm{I/II}}^2\over\d k^2}\right|_{k\to0} &=& 
{1\over2}\left[\tr[C_0^2] +(1\pm1)\,\tr[\Delta C^2]\right]
= c_0^2 + {1\pm1\over2}\tr[\Delta C^2] \, ,\label{eq:varpi2b}
\end{eqnarray}
and
\begin{equation} \label{eq:varpi4b}
\left.{\d^2\omega_{\mathrm{I/II}}^2\over\d(k^2)^2}\right|_{k\to0} 
=
{\tr[Z^2]  \pm \tr[Z^2] \over 2}\pm\tr[C^2_0]\left(-\tr[Y^2]+
{\tr[\Delta C^2] \over \tr[\Omega^2] }\right) \, .  
\end{equation}
To achieve conditions $C1$ and $C2$ in the 2-component BEC the effective coupling between the hyperfine states has to vanish, $ \tilde{U}_{12}=0$. This can be implemented by imposing a particular transition rate $\lambda = -2 \sqrt{\rho_{10}\;\rho_{20}} \; U_{12}$.
In addition to the fine tuning of $\lambda$, the paramters ($U_{ii},\rho_{i0},m_{i}$) have to be chosen so that the speed of sound simplifies to:
\begin{equation}   \label{eq:c0av} \fl
c_0^2 
={\tilde U_{11} \;\rho_{10}\over m_1} 
=  {\tilde U_{22} \;\rho_{20}\over m_2} 
= \frac{m_2 \rho_{10} U_{11} + m_1 \rho_{20} U_{22} + U_{12} (\rho_{10} m_1 + \rho_{20} m_2) }{2 m_1 m_2 }.
\end{equation}
While one eigenfrequency always remains zero, $\omega_{0,\mathrm{I}}\equiv0$, for the second phonon mode we get
\begin{equation}
\omega_{0,II}^2 = \frac{4 U_{12} (\rho_{10} m_2 + \rho_{20} m_1) c_0^2}{ \hbar^2} \,.
\label{eq:om2}
\end{equation}
The mass of the modes are then defined as
\begin{equation} \label{mass_final}
m_{\mathrm{I/II}}^2 = \hbar^2 \omega_{0,\mathrm{I/II}}^2/c_0^4 \,,
\end{equation}
and thus the AM corresponds to one massless particle $m_{\mathrm{I}}=0$ and one massive particle,
\begin{equation} \fl
m_{II}^2 = {
8 U_{12} (\rho_{10} m_1+\rho_{20} m_2) m_1 m_2 
\over
          [m_2 \rho_{10} U_{11} + m_1 \rho_{20} U_{22} 
         + U_{12} (\rho_{10} m_1 + \rho_{20} m_2)]
} \approx  {m^2  \;
8 U_{12}
\over [U_{11}+2U_{12}+U_{22}]
}
\end{equation}
propagating in the acoustic Minkowski space-time in the hydrodynamic limit. For higher wave numbers we obtain LIV in the form of equation (\ref{ext_disp_rel}), and the coefficients $\eta_{2}$ and $\eta_{4}$ for the two modes are: 
\begin{eqnarray} 
&\fl \eta_{2,\mathrm{I/II}}
=\frac{\hbar^2}{4 c^4_0} \; 
\frac{\rho_{10} m_1 + \rho_{20} m_2}{\rho_{10} m_2 + \rho_{20} m_1} \;
\frac{\omega_{0,\mathrm{I/II}}^2 }{m_1m_2}
\approx\left(\frac{m_{\mathrm{I/II}}}{M_{\rm eff}}\right)^2= 
\left( \frac{ \mathrm{mass}_{\,\mathrm{quasiparticles}} } {\mathrm{Planck\;scale}_{\,\mathrm{effective}}} \right)^2 \, ;
\label{eta2_final}
\\
&\fl \eta_{4,\mathrm{I/II}} =  \frac{1}{4}
\left[
\frac{\gamma_{\mathrm{I/II}}\,m_{1}\rho_{10}+\gamma_{\mathrm{I/II}}^{-1}\,m_{2}\rho_{20}}{m_{1}\rho_{20}+m_{2}\rho_{10}}
\right] \, , \label{eta4_final}
\end{eqnarray}
where $\gamma_{\mathrm{I}}=1$ and $\gamma_{\mathrm{II}}=m_{1}/m_{2}$ are dimensionless coefficients, and $M_{\mathrm{eff}}=\sqrt{m_{1} m_{2}}$ is defined as the analogue Planck mass. 

From the expression it is clear, that the quadratic coefficients (\ref{eta2_final}) are non-universal. While one is always zero, $\eta_{2,\mathrm{I}}\equiv0$, the second $\eta_{2,\mathrm{II}}$ depends on the interaction constant $U_{12}$. For $U_{12} \ll (U_{11}+U_{22})$, which is equivalent to $m_{\mathrm{II}} \ll \sqrt{m_{1}m_{2}}$, so $\eta_{2,\mathrm{II}}$ is suppressed. 
However, there is no further suppression --- after having pulled out a factor $(\hbar / M_{\mathrm{Pl}})^2$ --- for the quartic coefficients $\eta_{4,\mathrm{I/II}}$. These coefficients are of order one and non-universal, (though they can be forced to be universal, for example if $\gamma_{\mathrm{I}} = \gamma_{\mathrm{II}}$ and the underlying bosons have equal masses $m_{1}=m_{2}$).  

The suppression of $\eta_2$, combined with the non-suppression of $\eta_4$, is precisely the statement that the ``naturalness problem'' does not arise in the current model.

%
\section{Summary and discussion}
We have presented an AM that can be used as a model for and motivation for conjectures in QGP. Low energy perturbations in a 2-component BEC with laser-induced transitions between the single-particle states reproduce both massive and massless quasi-particles in an emergent space-time. Furthermore, we have investigated higher energy perturbations for a uniform condensate without background flow. This corresponds to particles propagating in Minkowski flat space-time with large momentum. While in the hydrodynamic regime the dispersion relation is LI, beyond it the dispersion relation has to be modified in a Lorentz violating way. 

Due to parity of the GPE we only obtained terms with even exponents of $k$. We calculated the quadratic and quartic dimensionless coefficients $\eta_{2,\mathrm{I/II}}$ and $\eta_{4,\mathrm{I/II}}$. A key observation is that the present model does not suffer from the  naturalness problem,  because the quadratic corrections are Planck suppressed, while at the same time the quartic coefficients $\eta_{4.\mathrm{I/II}}$ have no further suppression, and are actually of order unity. 


\end{document}